\documentclass[utf8]{article}
\usepackage{latexsym,amssymb}
\usepackage{amsmath}
\usepackage{wasysym,stmaryrd}
\usepackage{xcolor,graphicx}
\usepackage{hyperref}
\usepackage{float} 
\usepackage{subfloat} 
\usepackage{babel}

\usepackage{wrapfig}
\usepackage[all]{xy}
\input{xy} \xyoption{all}

\newcommand{\CSforF}{CS4F}

\begin{document}

\title{Computer Science for Future\\Sustainability and Climate Protection in the Computer Science Courses of the HAW Hamburg
}

\author{Elina Eickstädt \and Martin Becke \and Martin Kohler  
        \and Julia Padberg\footnote{email: ComputerScience4Future@haw-hamburg.de }
				\\[3mm]
				\large Hamburg University of Applied Sciences }

\maketitle
\begin{abstract}
Computer Science for Future (\CSforF) is an initiative in the Department of Computer Science at HAW Hamburg. The aim of the initiative is a paradigm shift in the discipline of computer science, thus establishing sustainability goals as a primary leitmotif for teaching and research. The focus is on teaching since the most promising multipliers are the students of a university. The change in teaching influences our research, the transfer to business and civil society as well as the change in our own institution. In this article, we present the initiative \CSforF~ and reflect primarily on the role of students as amplifiers in the transformation process of computer science.
\end{abstract}
\tableofcontents

\section{Computer Science is Cross-Sectional}
The scientific discipline of computer science has played a steadily growing role in shaping society over the past decade.
Computer science defines processes in companies, increasingly models modes of interaction in society and determines how services and goods are handled. 
In an increasingly granular networked and automated world, the infrastructure of our coexistence is determined by decisions made in computer science. 

In short, computer science has affected, if not already permeated, almost every area of our existence. The success of the digital technologies in recent decades has been strongly influenced by ideas and concepts of the Silicon Valley, which, with mantras like “move fast and break things”, have shaped an understanding of computer science as a disruptive technology supplier. Slogans such as “make the world better” were often misunderstood in simply doing it differently.

This attitude is increasingly being challenged, and there is a growing recognition that digitisation and informatics are systematically integrated into the
Global sustainability should be at the service of global sustainability.
The relationship between informatics and sustainability is also the focus of various science policy organisations (e. g. %
the 
"`Gesellschaft für Informatik"'%
    \footnote{\href{https://informatik2021.gi.de/informatik2021}{https://informatik2021.gi.de/informatik2021} (visited 2. 1. 2023)} (Society for Informatics),
"`Bits und Bäume"'%
        \footnote{\href{https://fahrplan22.bits-und-baeume.org/bitsundbaeume/}{https://fahrplan22.bits-und-baeume.org/bitsundbaeume/} (visited 2. 1. 2023)} (Bits and Trees),
FIfF\footnote{\href{https://www. fiff. de/}{Forum Computer Scientists for Peace and Social Responsibility} (visited 2. 1. 2023) },%
ISOC.DE\footnote{\href{https://isoc. de/}{Internet Society, German Chapter} (visited 2. 1. 2023) },%
netzpolitik.org\footnote{\href{https://netzpolitik. org/}{netzpolitik. org} (visited 2. 1. 2023) } or within the 
UN\footnote{Promotion and protection of all human rights, civil, political, economic, social and cultural rights, including the right to development \href{https://www. article19. org/data/files/Internet_Statement_Adopted. pdf}{ (A/HRC/32/L. 20) } (visited 2. 1. 2023)}).

This change can also be seen in research, for example, the number of research publications on net zero emissions targets increased to 1. 6 million between 2001 and 2020, representing 4.9\% of global publications compared to 1.2\% in 2001 \cite{elsevier_report_pathways}.

Although there are already various approaches to establish sustainability aspects in higher computer science education, e.g.~\cite{Vakaliuk_et_al2020,Fucci_et_al_2017,Fernandes_et_al_2022,Abdillah_et_al_2018,Smith_et_al_2022} these questions are still far from being established in teaching.
It must be questioned whether today’s computer scientists have sufficient knowledge to consider other targets such as efficiency, optimization or innovation.
In the context of current global challenges such as climate change, decisions in information systems can have far-reaching consequences. For example, the selection of the transaction technology can have a significant impact on the CO$_2$ fingerprint \cite{MIGLANI2020395}.

These costs are rarely part of the requirements process because there is little or no knowledge about the  dependencies that can be expected. 
Much depends on the use to which the technology is optimized and the extent to which users and producers are aware of its influence. 
The avoidance of greenhouse gases in the context of climate change is only a very small part of what can be achieved with new objectives, new projects and new tools. 
Teaching is the our main focus, but its content  hast to  be kept up to date basically through the connection to research.

\section{\CSforF~ Experiments}
\label{s. motif}
In the Department of Computer Science at the Hamburg University of Applied Sciences (HAW), we would like to react to the demands of our time and have launched initiatives at the beginning of 2022 that have found a conceptual superstructure under the title Computer Science for Future (\CSforF). Through \CSforF, the Department of Computer Science would like to initiate a paradigm shift with its teachers and researchers, with the staff and students, and not least with the existing laboratories. 

 With \CSforF, we are aiming for a paradigm shift in the discipline of computer science by establishing sustainability goals as the primary guiding principle for teaching and research.
In order to address these sustainability goals with diverse methodologies, in heterogeneous groupings and with different objectives, we want to conceive of concrete strategies, measures and activities as experiments that are subject to permanent feedback and gain our insights in small, tangible experiments. 
We accept our responsibility and focus on the future viability of computer science. That is why \CSforF~ launches a transformation process to address the imperative of climate protection and sustainability from the perspective of informatics:
 
\begin{description}
\item[enhancement of the curriculum:]\hfill\\
In teaching, new courses are to be developed as well as existing ones adapted to meet the requirements of the \CSforF.
\item[permanent integration into research and organization:]\hfill \\
   It is important to create reward systems for the respective institution, as well as to establish structures that consolidate the orientation of \CSforF~ in the long term.
\item[communication with others:]\hfill\\
The aim is to offer points of exchange and linkages in order to set in motion a multilateral process of problem awareness and action awareness.
\end{description}
The process of change will affect different aspects on a large and small scale and thus take place at different levels of granularity. In particular, \CSforF~ derives its focus from the Sustainable Development Goals set by the United Nations, also known as the Sustainable Development Goals (SDGs).

\begin{figure}[htbp]
\centering
\includegraphics[width=0.95\textwidth]{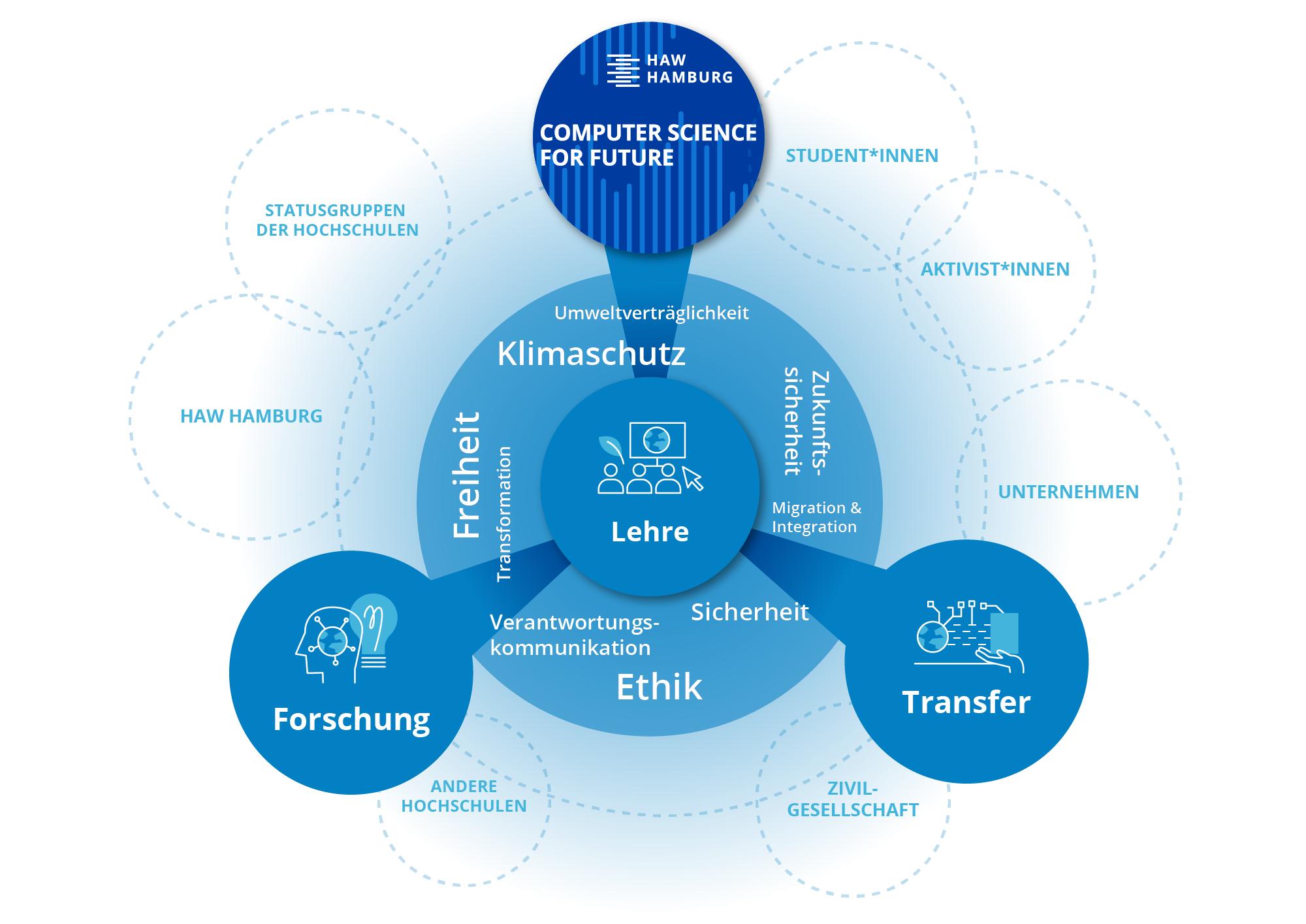}
\caption{Effect of \CSforF}
\label{fig:Effect_CS4F}
\end{figure}

However, \CSforF~ does not only see itself as responsible for minimising damage, but also wants to make clear to all designers of and participants in information technologies the high level of responsibility that goes hand in hand with their use and offer solutions to meet this responsibility.  Strategically, \CSforF~ thus pursues various goals, which, as illustrated in Figure \ref{fig:Effect_CS4F}, unfold the desired effect in the interaction of different sub-areas:
 teaching, research, transfer and civil society initiatives. Thus, the \CSforF~ initiative is also revitalised by the students who have pursued a different educational focus. The figure illustrates that teaching is the focus and triggers further work in the other sub-areas of research, transfer and civil society initiatives. 

For research in particular, it is to be expected that research tasks will result from questions in teaching. Teaching must result from established scientific foundations that can be addressed in the context of research on sustainability topics in the department. Currently, the teaching material for wide areas of sustainability and computer science can be considered insufficient.

In the area of transfer, the teaching has a direct effect firstly through our graduates, who take the knowledge of connections between computer science and sustainability with them to their future jobs. In addition, we strive for cooperations within the framework of the interdisciplinary laboratory "Creative Space For Technical Innovations" (CSTI)\footnote{\href{https://csti.haw-hamburg.de/}{https://csti.haw-hamburg.de/}(last visited 6.6.2022)}, which result from approaches to solutions from research and teaching.

But teaching is at the heart of the changes. One of the main goals of \CSforF~ is to strengthen the motivation and problem awareness of students, the future developers. We see the students at the universities as multipliers in order to achieve sustainable computer science in the long term.  

The transformation process in the Department of Computer Science driven by \CSforF~ is supported by all those who participate, i.e. students, staff and professors, who are involved in the following areas 
ideas of sustainability (from the perspective of computer science), especially climate protection.
\CSforF~ thus serves the 
representation, information and coordination in order to develop
to develop visions and goals, to implement and support measures and to promote continuity and funding.
The activation of students is a central component of the initiative in order to generate activity from the intrinsic motivation. \CSforF~ builds on the demands and initiatives of Fridays for Future or Students for Future and provides already active students with a framework in which ideas for implementation in computer science can be scientifically expressed and discussed.
Furthermore, students who have not been involved so far should be informed about the effects of their decisions and involved in the discussion. For \CSforF, strengthening self-responsibility is an essential key to establishing new goals, i. e. in the creation, operation and safeguarding of software and hardware.

\section{Teaching as a Multiplier}
\label{s. tehr}
With Computer Science for Future (\CSforF), the Department of Computer Science aims to become sustainable in the sense of the UN's Sustainable Development Goals. This target is demanding: very comprehensive, very complex and also very controversial. Therefore, its implementation requires an adaptable transformation process that affects teaching, research and our infrastructure. Teaching has a special role to play here, because our students represent both the greatest innovative potential and the greatest multiplier. Both forces can be used in our teaching and integrated into the transformation.
The teachers see themselves as learning coaches in the sense of \cite{krause-steger_integration_2019} and focus their teaching more on education for sustainable development as well as on the concrete social challenges and on interdisciplinarity in teaching. 

\subsection*{Research Learning}
In \cite{lingenau_integration_2019} it is stated that digitisation as well as education for sustainable development in the field of higher education needs such teaching/learning formats to provide students with those competences that are needed to solve problems in the future that are currently not yet known. The concept of inquiry learning seems suitable to combine both directions: 
\begin{itemize}
	\item collaborative learning processes include learning and knowledge achieved in non-hierarchical groups, as well as. 
\item dealing with complexity and uncertainties and reflective questioning also aim at action competences of sustainable development.

\end{itemize}

The experience of the last two years of the pandemic has clearly shown that collaborative learning under poor conditions can lead to more stress among students. \CSforF~ will critically reflect on how to design the curriculum to give students space to think and time to engage with sustainability issues. This also applies to the implementation of collaborative learning in the true sense. This requires a constant exchange with students as well as a flexible adaptation of teaching. Existing structures must be strengthened and expanded. 
The members of the interdisciplinary laboratory "Creative Space For Technical Innovations" (CSTI) are contact persons for research, teaching and transfer and, due to the experience in research-based teaching already available in the laboratory, an important initial point of reference for the desired transformation.  

\begin{figure}
	\centering
		\includegraphics[width=0.60\textwidth]{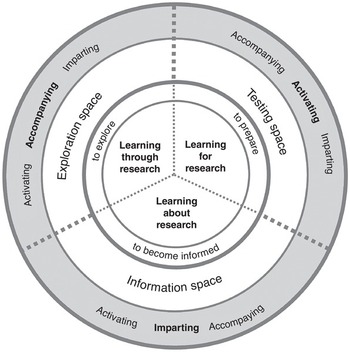}
	\label{fig:cambridgeResearchLearning}
\caption{Research Learning}
\end{figure}

In \cite{mieg_ambos_brew_galli_lehmann_2022} a suitable classification of the possibilities of implementation is given (see Figure  \ref{fig:cambridgeResearchLearning}). It is particularly profitable that different target groups, also with different objectives and backgrounds, can be addressed.

\subsection*{Transformative Science}

The serious turn towards integrating the UN's Sustainable Development Goals (SDGs) and accepting the societal role that informatics plays in shaping the world and its relationships also means transforming what we understand as science principles. 
The German Advisory Council on Global Change (WBGU)
made this clear programmatically as early as 2011 with the 
catchphrase "transformative research"\cite{wissenschaftlicher_beirat_der_bundesregierung_globale_umweltveranderungen_welt_2011}. In the wake of global crises and the need for rapid change, knowledge gain and theory building must themselves become part of the transformation. The aim of transformation research is to describe, explain, evaluate and support transformations towards a sustainable society. A transformative science supports and is itself part of such transformation processes and can thus contribute to change in a very concrete way\cite{schneidewind_vom_2015}. The advantage is also that science can and must change itself under the impression of the examined processes, as it also offers a deeper understanding and possibility to examine inner regularities and processes as part of the transformation movement. Transformative questions can influence the processes (what changes within transformations), the dynamics of change (how do transformation processes take place) or the driving forces (through what / by whom are transformation processes supported). Sustainability transformations are open, non-linear processes characterised by high uncertainty of prediction \cite{schapke_linking_2017}. 
By embedding such processes, systemic thinking, concrete questions and transdisciplinary working methods are necessary. The real-world reference and the inclusion of different, also non-academic experts and voices favours research formats such as workshops, reallabs or field experiments. For computer science, this means a further strengthening of agile and incremental approaches. Beyond this, however, is the involvement of civil society and individual non-academic actors. 
The clarification of relevance (what is important for a successful transformation instead of what is important for the scientific discipline) benefits from cooperations in the sense of a co-design both for the relevant research questions and for the research goals. 
Especially for the clarification of relevance (what is important for a successful transformation instead of what is important for the scientific discipline), such cooperations in the sense of co-research and co-design of the research goals, but especially of the relevant research questions, are essential. 

With the \CSforF~ initiative, we want to give the emerging activities and initiatives an essential role in helping to clarify and develop what actually needs to be researched and transformed. Here, research can then emerge from teaching, with student initiatives becoming the questioners for research projects or with interest groups conducting non-academic research. 

\subsection*{Persistence in Student Projects}
At the heart of the activities around \CSforF~ is the future viability of computer science and the postulate that computer science students are the ones who drive innovative and implementable solutions. Individual projects become prototypical experiments that serve as concrete learning and practical experiences that can lead to long-term change \cite{munck_af_rosenschold_inducing_2019}. However, this is contrasted by the short length of time students stay in such activities due to workload and degree attainment.   \CSforF~ is therefore also an initially rudimentary platform to bundle and continue student activities.

\section{Ongoing Activities}
Since the beginning of the year, we have started these initiatives and are thus still in the first phase, the implementation of our trials. The focus of the first stage in particular is on two goals. One is to increase visibility internally and externally, and the other is to rebuild the curriculum within the framework of the existing examination regulations . This generally means creating communication platforms and new teaching offers to drive projects and inform about your progress:

\subsection*{Collaboration in \CSforF}
\label{s.orga}
The initiatives and actions for the internal projects are developed and communicated in this MS team \CSforF~ \textit{Computer Science for Future}. In this way, we have established a communication structure which, among other things, serves to
\begin{itemize}
    \item to publicise the current and planned measures,
    \item to develop new measures from suggestions and ideas,
    \item cooperation within the HAW and
    \item to organise the cooperation with others.	
\end{itemize}
In order to be able to respond to current requirements, the process is carried out in an agile manner, i.e. measures are regularly discussed throughout the department across all status groups, addressed and considered in retrospectives. This includes initiatives in teaching, in research and in the organisation of the Department of Computer Science.  

\subsection*{Podcast \CSforF}

Accompanying the implementation of the project is a podcast created by students professors and employees of the university. In each episode, the team talks to activists, experts and scientists about climate change mitigation, sustainability in the context of computer science. The podcast also features some basic discussions about ethical issues surrounding the impact of technology on society. The aim of the podcast is on the one hand to report on the work on the project, but also to give those interested in computer science an insight into the different facets and possibilities of sustainable computer science.
In the best case, an intrinsic motivation can be triggered to become active oneself. Of course, the podcast will also offer an interface to directly address young students and to get in touch with schools and companies. The goal is not to make the podcast primarily for insiders, but to introduce listeners to the topics. It also focuses less on persuasion and more on providing broad information about how computer science works and what impact computer science can have on society.  

\subsection*{Climate Orientation Unit}
 \CSforF~ has accompanied student activities from the context Students for Future and supported them with its own offers in the Climate Orientation Unit (Klima OE), a student event series. The Climate OE offer was cross-university and various events were held in one week to draw attention to the topic of sustainability, to show possible solutions and to discuss room for manoeuvre. \CSforF~ has formulated the position from within the department and fed it into the process. In this discussion, the enormous need for an interdisciplinary exchange once again became more than clear.  

\subsection*{Sustainability goals in courses}
Currently, elective courses are held on various sustainability topics. These cover the following topics, among others:
\begin{itemize}
	\item Relationships between human rights and protocols,
	\item Challenges and solutions to the climate crisis
	\item climate and information technology 
	\item ethical responsibility of computer scientists
	\item machine ethics
	\item environmental informatics
\end{itemize}

The relationship between human rights and protocols is relevant to our teaching.
protocols are relevant to our teaching, because protocols have an effective
influence on rights such as freedom of expression or the right of assembly.
right of assembly. The lecture series \textit{Our House Is On Fire}
by the AstA of the University of HH and Fridays for Future is already taking place for the third semester and focuses on the challenges and solutions of the climate crisis.
It is offered as an elective course in the bachelor's degree programmes in computer science. Another course focuses on climate and computer science and examines the challenges in IT in the area of tension between the environment, technology and the economy. One course focuses on the responsibility of computer scientists and requires fundamental reflections on computer science and its use. Another deals with machine ethics, i.e. the question of what it means when machines are increasingly thought of and constructed more autonomously and intelligently, so that their activities become increasingly ethically relevant. There is also a seminar on environmental informatics, which deals with the generation and processing of environmental data as well as environmental processes.

In addition, we have started to introduce various CS4F topics in three mandatory courses:
\begin{itemize}
	\item 
In the first semester, the course "Fundamentals of Computer Engineering" was expanded to include interdisciplinary concepts. In particular, concepts from sociology were used to illustrate the contrast between solutionism and technology criticism. In addition, a basic overview of ethics was given and the role of ethics in computer science was discussed. This included an introduction to various ethical theories, such as contractualism, autonomy orientation, and John Rawls' modern contract theory, and how these concepts relate to computer science.
\item 
The course "`Fundamentals of Computer Science"' takes place in the first semester. The lecture has been extended to include the topic of sustainable software development. Sustainable software development refers to the creation of software that is sustainable and resource-efficient in the long term and considers not only technical aspects, but also social, ecological and economic factors. CO$_2$ emissions associated with the use of ICT systems are a significant contributor to global greenhouse gas emissions, and addressing these emissions is an important part of efforts to combat climate change. By implementing CO$_2$-conscious computing practices, businesses, governments, and individuals can help reduce their greenhouse gas emissions and contribute to a more sustainable future. 
\item 
In the third semester in the "Applied Computer Science" program, graph theory and graph algorithms are taught. These algorithms are used to analyzeanalyse and optimize complex systems in areas such as transportation, energy and communication. The lab will provide practice in implementing such algorithms, but students will also learn to make their design decisions consciously and communicate them appropriately. This semester, this approach has been extended to make students aware of the potential socially relevant implications of their own implementation. They are expected to learn about the potential impact of their own work on society and to use ethical guidelines from the "Gesellschaft für Informatik" (German Informatics Society) to address ethical decisions and be aware of the potential consequences. Students should also learn to communicate their work and findings in an understandable way and to participate in societal discussions.
\end{itemize}
These cases are some of the learning formats developed or adapted under the CS4F paradigm at the HAW. The CS4F framework is conceptualized as an open platform for students and teachers alike to create and experiment with new course modules, projects and learning formats that integrate sustainability issues, transdisciplinary and cross-disciplinary thinking and concrete projects tackling sustainability issues in or by the use of computer science. 

The first results are promising and can develop into a solid path how to integrate sustainability issues in the education of future computer scientists and thus impacting the way these students act as professionals in their respective fields. 

\bibliography{CS4F_BB22} 
\bibliographystyle{plain}
\end{document}